\documentclass{moriond}

\usepackage{amsfonts} 
\usepackage{amsmath} 
\usepackage{amssymb} 
\usepackage{bm} 
\usepackage{xcolor}
\usepackage{cleveref}
\usepackage[titletoc,title]{appendix}
\usepackage{enumerate}
\usepackage{slashed}

\newcommand{\al}{\alpha}
\newcommand{\bt}{\beta}

\newcommand{\la}{\lambda}

\newcommand{\ksq}{\ka^{\prime 2}}
\newcommand{\ka}{\kappa}

\newcommand{\vL}{\ensuremath{\mathcal{L}}}

\newcommand{\Or}{\mathcal O}

\newcommand{\sq}{^{2}}

\newcommand{\dslash}[1]{#1 \llap{/\kern-0.5pt}}
\newcommand{\Dslash}[1]{#1 \llap{/\kern+1.2pt}}
\newcommand{\DDslash}[1]{#1 \llap{/\kern+2.3pt}}
\newcommand{\dslashh}[1]{#1 \llap{/\kern+1pt}}

\newcommand{\bea}{\begin{eqnarray}}
\newcommand{\eea}{\end{eqnarray}}
\newcommand{\bma}{\begin{pmatrix}}
\newcommand{\ema}{\end{pmatrix}}
\newcommand{\nn}{\nonumber}

\newcommand{\Photo}{}

\begin{document}
\vspace*{4cm}
\title{Testing left-right symmetric models\footnote{Talk presented at the $50$th Recontres de Moriond (EW session), La Thuile, March $20$, 2015.} }

\author{ W. Dekens}

\address{Van Swinderen Institute for Particle Physics and Gravity,\\ University of Groningen, Nijenborgh 4,\\
9747 AG Groningen, The Netherlands}

\maketitle\abstracts{
The difference between left- and right-handed particles is perhaps one of the most puzzling aspects of the Standard Model (SM). In left-right models (LRMs) the symmetry between left- and right-handed particles can be restored at high energy. Due to this symmetry these models are quite predictive with regards to experimental observables, making them interesting beyond the SM candidates. Here we discuss the more symmetric LRMs, the experimental constraints, and the fine-tuning  present in the Higgs sector.
}

\section{Introduction}
Left-right (LR) models \cite{LRSM,Mohapatra:1974hk,Senjanovic1975,Senjanovic1979,Deshpande:1990ip} extend the SM gauge-group to $SU(3)_c\times SU(2)_L\times SU(2)_R\times U(1)_{B-L}$. These models allow one to interpret the $U(1)$ generator in terms of baryon and lepton number and naturally incorporate the see-saw mechanism for neutrino masses \cite{Mohapatra:1979ia,Mohapatra:1980yp}. Furthermore, in some grand unified theories (GUTs), such as $SO(10)$ and $E_6$, the gauge group of the LRM can appear as an intermediate step \cite{Rizzo:1981su}. Perhaps the most attractive feature of LRMs is the possibility of a symmetry  between left- and right-handed particles at high energies. Such LR symmetric models are invariant under  parity ($P$) and/or charge conjugation ($C$) at high energies. Thus, $P$- ($C$-)symmetric LRMs account for the asymmetry between left and right in the SM by spontaneous breaking of $P$ ($C$).


Here we focus on the more symmetric LRMs which might be the most attractive from a theoretical standpoint. We introduce the \textit{minimal} LRM and present the most general Higgs potential in the next section. Experimental constraints, especially those from $B$- and $K$-meson mixing, and the fine-tuning in the Higgs sector are discussed in section \ref{PorC}.

\section{Minimal left-right models}\label{introLR}
The gauge group of left-right models is given by $SU(2)_L \times SU(2)_R \times U(1)_{B-L}$. The fermions are assigned to representations of this gauge group as follows,
\bea Q_L &=&\bma u_L\\d_L\ema \in (2,1,1/3), \qquad Q_R = \bma u_R\\d_R\ema\in (1,2,1/3),\nn\\
L_L &=&\bma \nu_L\\l_L\ema \in (2,1,-1), \qquad L_R = \bma \nu_R\\l_R\ema\in (1,2,-1).\eea
Given the above representations, a scalar bidoublet, $\phi\in (2,2^*,0)$, is required in order to allow for fermion mass terms.  
In addition, the \textit{minimal} LRM (mLRM) \cite{Mohapatra:1979ia,Duka:1999uc,Deshpande:1990ip,Zhang:2007da} introduces two scalar triplets $\Delta_{L,R}$ assigned to $(3,1,2)$ and $(1,3,2)$, respectively
\bea \phi = \bma \phi_1^0 & \phi_1^+\\ \phi_2^- & \phi_2^0 \ema ,\qquad
\Delta_{L,R} = \bma \delta^+_{L,R}/\sqrt{2} & \delta^{++}_{L,R} \\ \delta^0_{L,R} & -\delta^+_{L,R}/\sqrt{2} \ema .
\label{scalars}\eea
These scalar fields acquire the following vacuum expectation values (vevs)
\bea\label{vevs}
& \langle \Delta_{R} \rangle = \sqrt{1/2}\bma 0&0\\v_{R}&0\ema ,\qquad \langle \phi \rangle =\sqrt{1/2} \bma \kappa &0\\0&\kappa' e^{i\al} \ema ,\qquad \langle \Delta_{L} \rangle = \sqrt{1/2}\bma 0&0\\v_{L}e^{i\theta_L}&0\ema .
\eea
The vev of the right-handed triplet field, $v_R$, defines the high scale of the LRM and breaks its gauge group down to that of the SM.
Instead, $\ka$ and $\ka '$ are responsible for electroweak symmetry breaking, $\sqrt{\ka\sq+\ksq}=v\simeq 246 \, \text{GeV}$. Finally, the vev of the left-handed triplet contributes to the Majorana masses of the neutrinos, such that one would expect it not to exceed the neutrino-mass scale by much, $v_L\lesssim 1\, {\rm eV}$. These vevs are determined by the conditions for a minimum of the Higgs potential, the most general form of this potential is
\bea
V_H&=&
-\mu_1\sq \,\text{Tr}(\phi^\dagger \phi) -\mu_2\sq \text{Tr}(\tilde\phi^\dagger \phi)-\mu_2^{*2}\text{Tr}(\phi^\dagger\tilde \phi) - \mu_{3L}\sq \text{Tr}(\Delta_L\Delta_L^\dagger)-\mu_{3R}\sq\text{Tr}(\Delta_R\Delta_R^\dagger)\nn\\
&&+\lambda_1 \,\big[\text{Tr}(\phi^\dagger \phi)\big]\sq
+\big[\la_2\text{Tr}(\tilde\phi^\dagger \phi)\big]\sq+\la_2^*\big[\text{Tr}(\phi^\dagger\tilde \phi)\big]\sq+\la_3 \, \text{Tr}(\tilde\phi^\dagger \phi)\,\text{Tr}(\phi^\dagger\tilde \phi)\nn\\
&&+\text{Tr}(\phi^\dagger \phi)\big[\la_4\text{Tr}(\tilde\phi^\dagger \phi)+\la_4^*\text{Tr}(\phi^\dagger\tilde \phi)\big]+\rho_{1L}\big[\text{Tr}(\Delta_L\Delta_L^\dagger)\big]\sq+\rho_{1R}\big[\text{Tr}(\Delta_R\Delta_R^\dagger)\big]\sq\nn\\
&&+ \rho_{2L}\text{Tr}(\Delta_L\Delta_L)\text{Tr}(\Delta_L^\dagger\Delta_L^\dagger)+\rho_{2R}\text{Tr}(\Delta_R\Delta_R)\text{Tr}(\Delta_R^\dagger\Delta_R^\dagger)+\rho_3\text{Tr}(\Delta_L\Delta_L^\dagger)\text{Tr}(\Delta_R\Delta_R^\dagger)\nn\\
&&+\rho_{4L}\text{Tr}(\Delta_L\Delta_L)\text{Tr}(\Delta_R^\dagger\Delta_R^\dagger)+\rho_{4R}\text{Tr}(\Delta_R\Delta_R)\text{Tr}(\Delta_L^\dagger\Delta_L^\dagger)\nn\\
&&+ \,\text{Tr}(\phi^\dagger \phi)\big[\al_{1L}\text{Tr}(\Delta_L\Delta_L^\dagger)+\al_{1R}\text{Tr}(\Delta_R\Delta_R^\dagger)\big]\nn\\
&&+\big(\text{Tr}(\tilde\phi^\dagger \phi)\big[\al_{2R}\text{Tr}(\Delta_R\Delta_R^\dagger)+\al_{2L}\text{Tr}(\Delta_L\Delta_L^\dagger)\big]+\text{h.c.}\big)\nn\\
&&+\al_{3L}\text{Tr}(\phi\phi^\dagger \Delta_L\Delta_L^\dagger)+\al_{3R}\text{Tr}(\phi^\dagger \phi\Delta_R\Delta_R^\dagger)+\bt_1\text{Tr}(\phi \Delta_R \phi^\dagger\Delta_L^\dagger)+\bt_1^*\text{Tr}(\phi^\dagger \Delta_L \phi\Delta_R^\dagger)\nn\\
&&+\bt_2\text{Tr}(\tilde\phi \Delta_R \phi^\dagger\Delta_L^\dagger)+\bt_2^*\text{Tr}(\tilde\phi^\dagger \Delta_L \phi\Delta_R^\dagger) +\bt_3\text{Tr}(\phi \Delta_R \tilde\phi^\dagger\Delta_L^\dagger)+\bt_3^*\text{Tr}(\phi^\dagger \Delta_L \tilde\phi\Delta_R^\dagger),
\label{HiggspotP}\eea
where all parameters apart from $\mu_2$, $\la_{2,4}$, $\al_{2L,2R}$, and $\bt_i$ are real.
Large amounts of fine-tuning result from the fact that the minimum conditions relate the different vevs, and thereby widely varying scales, to one another. We will expand on this point in section \ref{PorC}.

One of the most characteristic ways in which LR models affect the observables we will consider, is through the interactions of the right-handed $W_R^\pm$ boson. The charged-current interactions involving the quarks, in the quark-mass basis, are given by 
\bea
\vL^{\text{CC}} = \frac{g_L}{\sqrt{2}}\overline U_L \gamma^\mu V_L D_L W_{L\mu}^+ + \frac{g_R}{\sqrt{2}}\overline U_R \gamma^\mu V_R D_R W_{R\mu}^+ +\text{h.c.}\,,
\eea
where $V_{L}$ and $V_R$ are the SM CKM matrix and its right-handed equivalent, and $g_{L,R}$ are the coupling constants of $SU(2)_{L,R}$. 
The right-handed current gives rise to additional contributions to $\bar K$-$K$ and $\bar B_{d,s}$-$B_{d,s}$ mixing observables, which for symmetric mLRMs are currently the best probes of the LR scale. These additional contributions depend on $V_R$, which, in turn, depends on the choice of LR symmetry. As a result, the constraints on LRSMs depend on the LR symmetry that is imposed.
 There are two possible transformations which qualify as symmetries between left and right
\bea
&P:& \qquad Q_{L}\longleftrightarrow\, Q_R , \,\quad \qquad\phi \longleftrightarrow\phi^{\dagger},\qquad \Delta_{L}\longleftrightarrow \Delta_{R},\nn\\
&C:&\qquad Q_{L}\longleftrightarrow \,(Q_R)^c , \qquad\phi\longleftrightarrow \phi^{T} ,\qquad \Delta_{L}\longleftrightarrow\Delta_{R}^*,
\label{C&Ptransf}\eea
where the superscript $c$ indicates charge conjugation. 
It turns out that the most symmetric option, $C$ \textit{and} $P$ invariance, is already excluded. There are multiple ways to implement both symmetries \cite{Ecker:1983hz,Ecker:1985vv}, but none of the possible models can simultaneously reproduce the observed $CP$ violation in $K$ mixing and the Belle and LHCb measurements of $CP$ violation in $B$ mixing ($\phi_d$), and give rise to a realistic Higgs spectrum \cite{Frere:1991db,Ball:1999mb,Ball:1999yi,Barenboim:2001vu}. This conclusion also holds for the minimal \textit{pseudomanisfest} \cite{Harari:1983gq,Langacker:1989xa} LRM, whose $P$-symmetric and real Yukawa couplings coincide with one of the $C$- and $P$-symmetric models.

\section{The $P$- \textit{or} $C$-symmetric LRMs}\label{PorC}
\begin{figure}[t!]
\centering
\includegraphics[width=75mm]{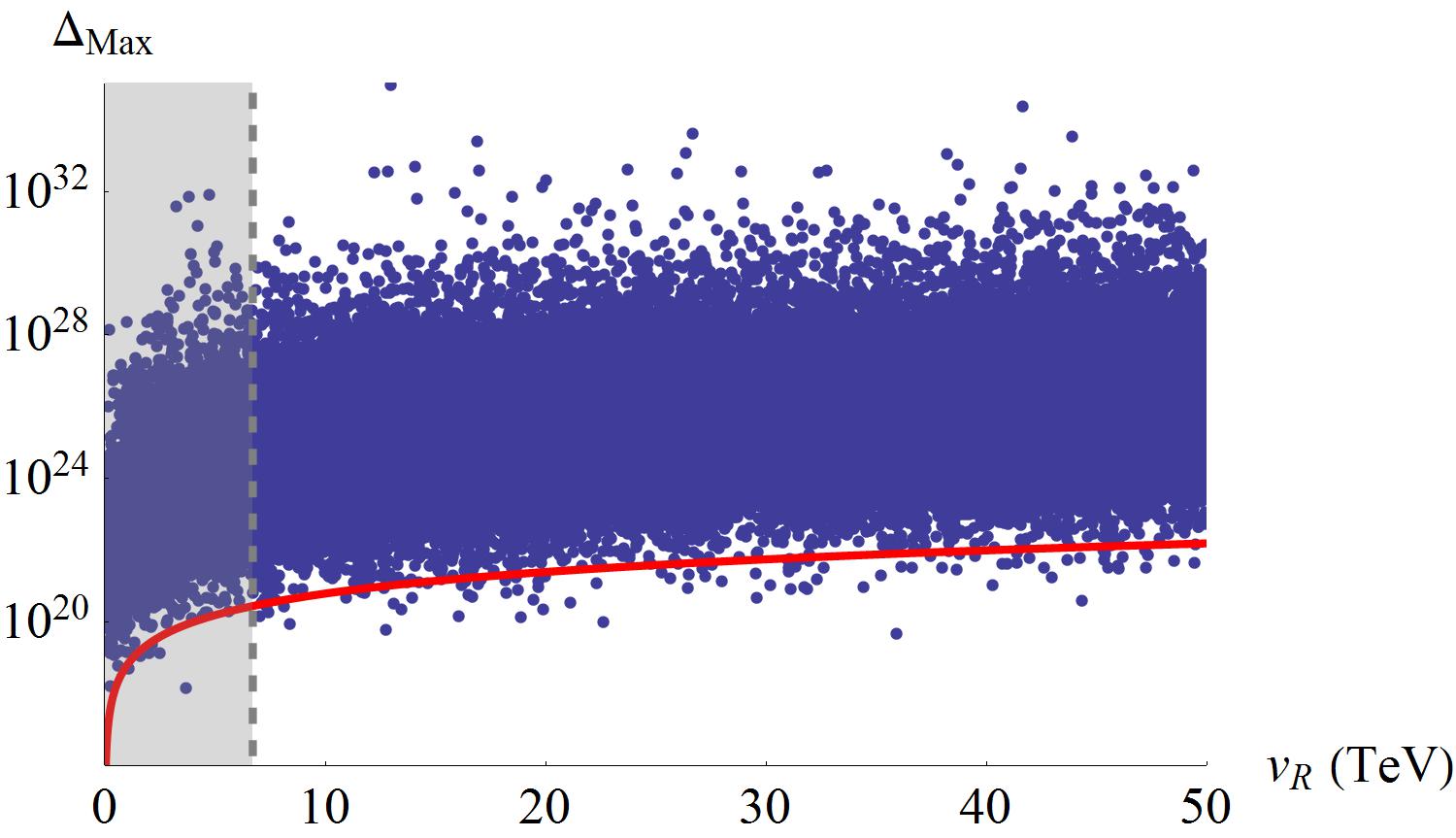} 
\includegraphics[width=75mm]{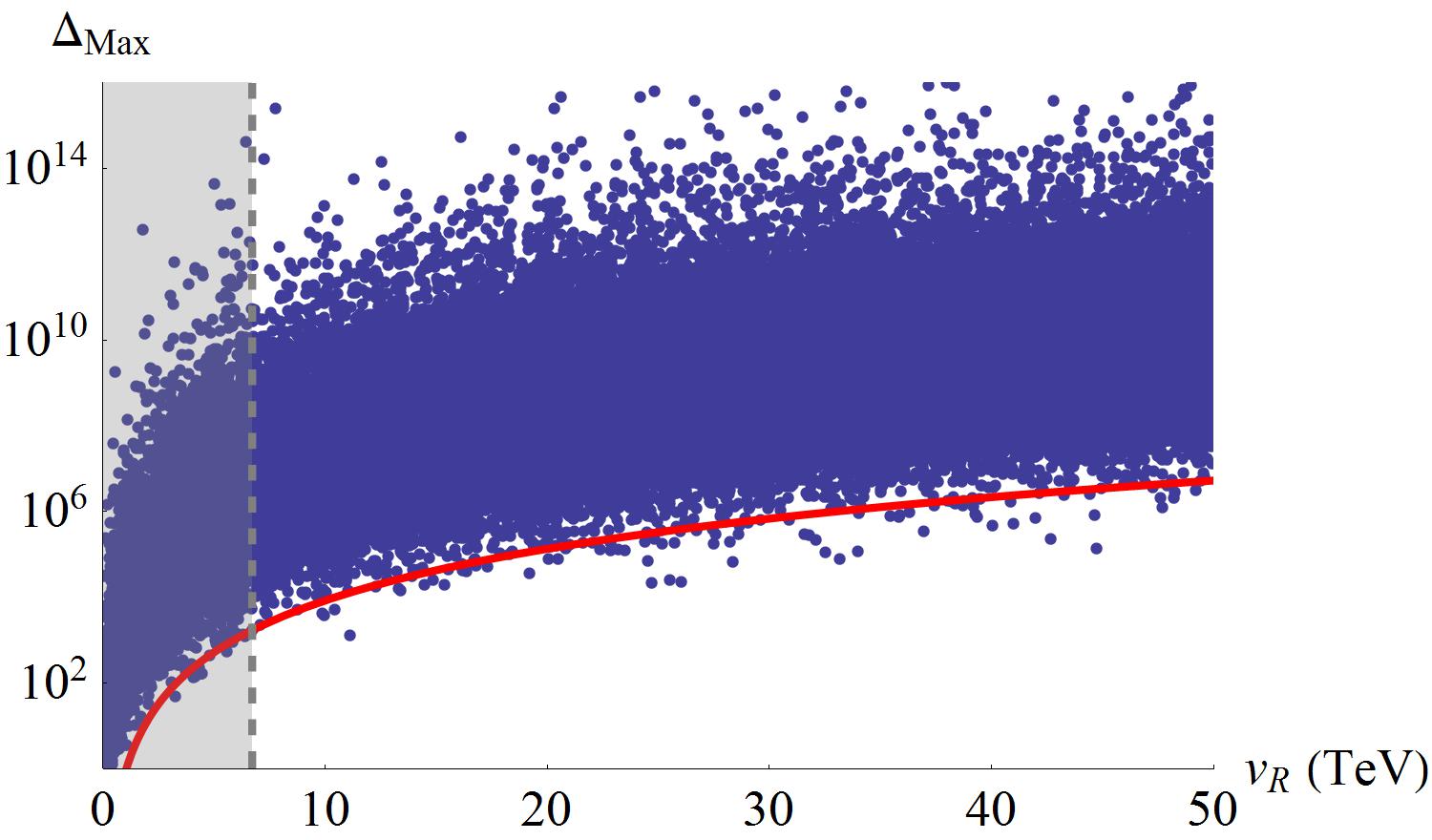} \\
$(a)$ \hspace*{7cm} $(b)$
\caption{The figure shows the fine-tuning measure $\Delta_{\text{Max}}$ as a function of $v_R$ in TeV for a $P$-symmetric $V_H$. The blue points are randomly generated points and the red line is chosen such that $0.1\%$ of the points are found below it. Figure \ref{Pfinetune}a shows $\Delta_{\rm Max}$ in the case where $\bt_i$ and $v_L$ are nonzero, while in Fig.\ \ref{Pfinetune}b we set $\bt_i=v_L=0$.} 
\label{Pfinetune}
\end{figure} 
In the $P$-invariant case the elements of the right-handed CKM matrix can be solved in terms of quark masses and $V_L$, such that it contains no free parameters. The exact solution was recently derived \cite{Senjanovic:2014pva,Senjanovic:2015yea}, which is approximately given by \cite{Zhang:2007da,Zhang:2007fn} $V^{ij}_L\simeq \pm V^{ij}_R$. Instead, in the $C$-symmetric case we have the relation, $V_R = K_u V_L^*K_d$, where  $K_{u,d}=\text{diag}(e^{i\theta_{u,d}},e^{i\theta_{c,s}},e^{i\theta_{t,b}})$ are diagonal matrices of phases. Although in the $C$-symmetric case $V_R$ contains additional free parameters, in both the $C$- and $P$-symmetric cases the combination of $B$- and $K$-mixing constraints result in a lower limit on the LR scale of roughly $M_{W_R}\gtrsim 3\, {\rm TeV}$ \cite{Bertolini:2014sua}. In the often considered case of the minimal \textit{manifest} \cite{Beg:1977ti,Langacker:1989xa} LRM, which has $P$-symmetric Yukawa couplings and $\al=0$, the lower bound is extended to $M_{W_R}\gtrsim 20\, {\rm TeV}$ \cite{Maiezza:2014ala} \footnote{
This limit also holds in the $P$-symmetric case if one does not incorporate a mechanism to set the QCD $\bar \theta$-term to zero. The neutron EDM limit then stringently constrains $\al$, effectively resulting in the manifest LRM \cite{Maiezza:2014ala}. In LRMs with a mechanism to enforce $\bar\theta=0$ the constraint on $\al$ is less severe, however, in this case the LRM predicts a relation between the EDMs of light nuclei \cite{Dekens:2014jka}, allowing for an experimental test of the model.}. This limit places the manifest scenario beyond the reach of future direct searches at the LHC or foreseen LHCb limits.
Although the current limits for the $C$- and $P$-symmetric cases are similar, $M_{W_R}\gtrsim 3\, \text{TeV}$ in each case, in principle, $B$- and $K$-mixing observables could distinguish between the two possibilities, as the two scenarios lead to different right-handed CKM matrices.

In contrast to the CKM matrices the $C$- and $P$-invariant Higgs potentials are rather similar \cite{Dekens:2014ina}. As a result, the fine-tuning that is required is very similar in either case and we focus on the $P$-invariant case here. We study this issue by solving the minimum equations for as many parameters, which we will denote by $p_i$, as there are equations. We then consider the dependence of these $p_i$ on the remaining parameters, $p_j$, through the fine-tuning measure, $\Delta$, often employed for supersymmetric models \cite{Ellis:1986yg,Barbieri:1987fn},
\bea
\Delta_i = \text{Max}_j\, \bigg|\frac{d \ln p_i}{d\ln p_j}\bigg|.
\label{FTdef}\eea
We calculate this fine-tuning measure for randomly generated points in parameter space, the results for the $P$-symmetric case are shown in Fig.~\ref{Pfinetune}, where we plot the maximum value of $\Delta_\text{Max}\equiv \text{Max} \,\Delta_i$ against $v_R$. The huge amount of fine-tuning, $\sim v_R\sq/v_L\sq$, in Fig.\ \ref{Pfinetune}a arises from the so-called `vev see-saw' relation \cite{Deshpande:1990ip}, $2\rho_1-\rho_3 \sim \bt_i\, \ka\ka'/v_L v_R$. This minimum condition calls for precise cancellations on the right-hand side if the $\rho_i$ parameters are to be $\Or(1)$. As has been noted \cite{Deshpande:1990ip} and can be seen from Fig.\ \ref{Pfinetune}b, the fine-tuning may be considerably decreased by setting $v_L=\bt_i=0$, see e.g.\ \cite{Zhang:2007da}. In this case, however, still a fine-tuning of order $\Delta = \Or(v_R^4/\ka_+^4)\gtrsim 100$ remains. We note that setting only $v_L$ to zero leads to the same reduction in the amount of fine-tuning \cite{Dekens:2014ina}. It remains to be seen whether these special cases can be justified or not \cite{Deshpande:1990ip,Barenboim:2001vu,Kiers:2005gh}.

\section{Conclusions}
In summary, LRMs with a LR symmetry are arguably the most attractive of the possible LRMs, but also the most constrained. The most symmetric models, invariant under $P$ and $C$, are already excluded by $B$- and $K$-mixing data. The LR scale of LRSMs with a $P$ or a $C$ symmetry is currently constrained by $B$- and $K$-mixing observables to be in the TeV range, $M_{W_R}\gtrsim 3\, \text{TeV}$, while future $B$-factory and LHCb data is expected to probe this scale up to roughly $8\, \text{TeV}$ \cite{Bertolini:2014sua}.
In the Higgs sector the potentials of the $P$-symmetric and $C$-symmetric  LRSMs turn out to be quite similar, and both require a  huge amount of fine-tuning, except in the case $v_L=0$.

\section*{Acknowledgments}
I thank Dani\"el Boer for collaboration on this work and reading of the manuscript. I thank the organizers of Moriond EW 2015 for financial support and the opportunity to present this work.

\end{document}